\documentclass[preprint,prc,showpacs,preprintnumbers,amsmath,amssymb]{revtex4-1}
\usepackage{graphicx}
\usepackage{dcolumn}
\usepackage{amsmath}
\usepackage{times}
\usepackage{mathtools}
\usepackage{rotating}

\begin{document}

\title{Series of broad resonances in atomic three-body systems }

\author{D.~Diaz}
\author{Z.~Papp}
\author{C.-Y.~Hu }
\affiliation{ Department of Physics and Astronomy,
California State University Long Beach, Long Beach, California, USA }

\date{\today}

\begin{abstract}
We re-examine the series of resonances found earlier in atomic three-body systems by solving the 
Faddeev-Merkuriev integral equations. 
These resonances are rather broad and line-up at each threshold with gradually increasing gaps, the same way 
for all thresholds and irrespective of the spatial symmetry. We relate these resonances to the Gailitis mechanism, which is a consequence of the polarization potential. 

\end{abstract}


\maketitle

\section{Introduction}

A couple of years ago we observed an accumulation of resonances above the thresholds \cite{prl,zsolti}. 
Other independent calculations have not confirmed these findings, only a few narrow resonances have been independently calculated.
This is understandable, since calculation of broad
resonances, especially in a multi-body system, is very complicated. The wave function of narrow resonances behaves very much
like a bound-state wave function, so they can be calculated by some slight modification of bound state techniques. 
The most common method is the complex rotation of coordinates. This technique renders the resonance state wave function 
to a square integrable one, and thus standard techniques  like variational expansion of the wave function become applicable. 
These methods, however, run into technical difficulties for broad resonances. Here, to uncover the resonances, they need 
a large rotation angle and the rotated continuum becomes more and more scattered making the identification 
of resonances increasingly difficult.

Our method is different. We start with the Faddeev integral equations with the modification proposed by Merkuriev \cite{fm-book}
and solve them by using the Coulomb-Sturmian potential separable expansion method  \cite{pz}.
Since the investigations of Refs.\ \cite{prl,zsolti} we improved the technique in Ref.\ \cite{k-m-p} making it more amenable to calculate
broad resonances. Therefore in Sec.\ II we outline our technique of solving the Faddeev-Merkuriev integral equations for resonant 
states while in Sec.\ III we show our results for $e-Ps$ resonances. Finally we summarize our findings and provide an explanation for 
the formation of the series of broad resonances.

\section{  Calculation of Coulomb three-body resonances }

\subsection{Faddeev-Merkuriev integral equations }

The  Hamiltonian of an atomic three-body system is given by
\begin{equation}
H=H^0 + v_1^C + v_2^C + v_3^C,
\label{H}
\end{equation}
where $H^0$ is the three-body kinetic energy 
operator and $v_\alpha^C$ denotes the Coulomb
interaction of each subsystem $\alpha=1,2,3$. 
We use the usual configuration-space Jacobi coordinates
 $x_\alpha$ and $y_\alpha$, where $x_\alpha$ is the distance
between the pair $(\beta,\gamma)$ and $y_\alpha$ is the
distance between the center of mass
of the pair $(\beta,\gamma)$ and the particle $\alpha$. The potential $v_\alpha^C$, the interaction of the
pair $(\beta,\gamma)$, appears as $v_\alpha^C (x_\alpha)$.
In an atomic three-body system two particles always have the same sign of charge. 
Without loss of generality, we can assume that they are particles $1$ and $2$, and thus $v_{3}^{C}$ is a 
repulsive Coulomb potential.

The wave function of a three-particle system is very complicated. It exhibits different  asymptotic behaviors reflecting the possible asymptotic fragmentations.  
In the Faddeev approach we split the wave function into components such that each component describes only one kind of asymptotic fragmentation \cite{fm-book}. 
The components satisfy a set of coupled equations, the Faddeev equations.

The  Hamiltonian (\ref{H}) is defined in the three-body 
Hilbert space. Therefore, the two-body potential operators are formally
embedded in the three-body Hilbert space,
\begin{equation}
v^C_\alpha = v^C_\alpha (x_\alpha) {\bf 1}_{y_\alpha},
\label{pot0}
\end{equation}
where ${\bf 1}_{y_\alpha}$ is a unit operator in the 
two-body Hilbert space associated with the $y_\alpha$ coordinate. 

The Coulomb potential  is a long range potential as it modifies the motion even at asymptotic distances. 
On the other hand, it also possesses some features of a short-range potential as it correlates the 
particles strongly and supports two-body bound states. These two properties are contradictory and require different treatments. 
In Merkuriev's approach  the three-body
configuration space is divided into different asymptotic regions \cite{merkur}. 
The two-body asymptotic region $\Omega_\alpha$ is
defined as a part of the three-body configuration space where
the conditions
\begin{equation}
(|x_\alpha|/  x_{0})^{\nu} <   |y_\alpha|/ y_{0},
\label{oma}
\end{equation}
with parameters $x_{0}>0$, $ y_{0} >0$ and $\nu > 2$ are satisfied. 
It has been shown that in $\Omega_{\alpha}$ the short-range character of the Coulomb potential prevails, 
while in the complementary region the long-range character of the Coulomb potential becomes dominant. 

\begin{figure}
\includegraphics[width=0.5\textwidth]{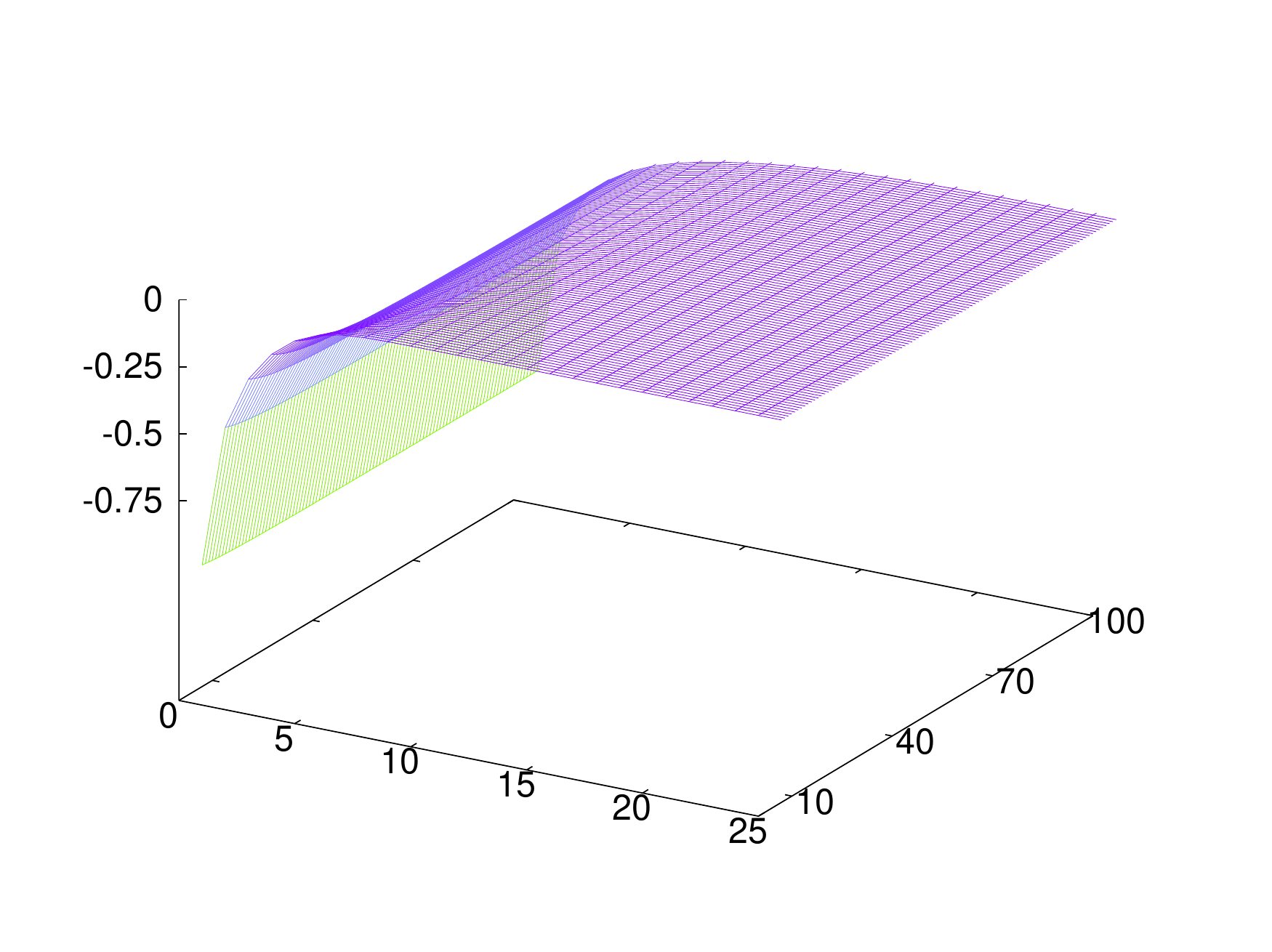}\includegraphics[width=0.5\textwidth]{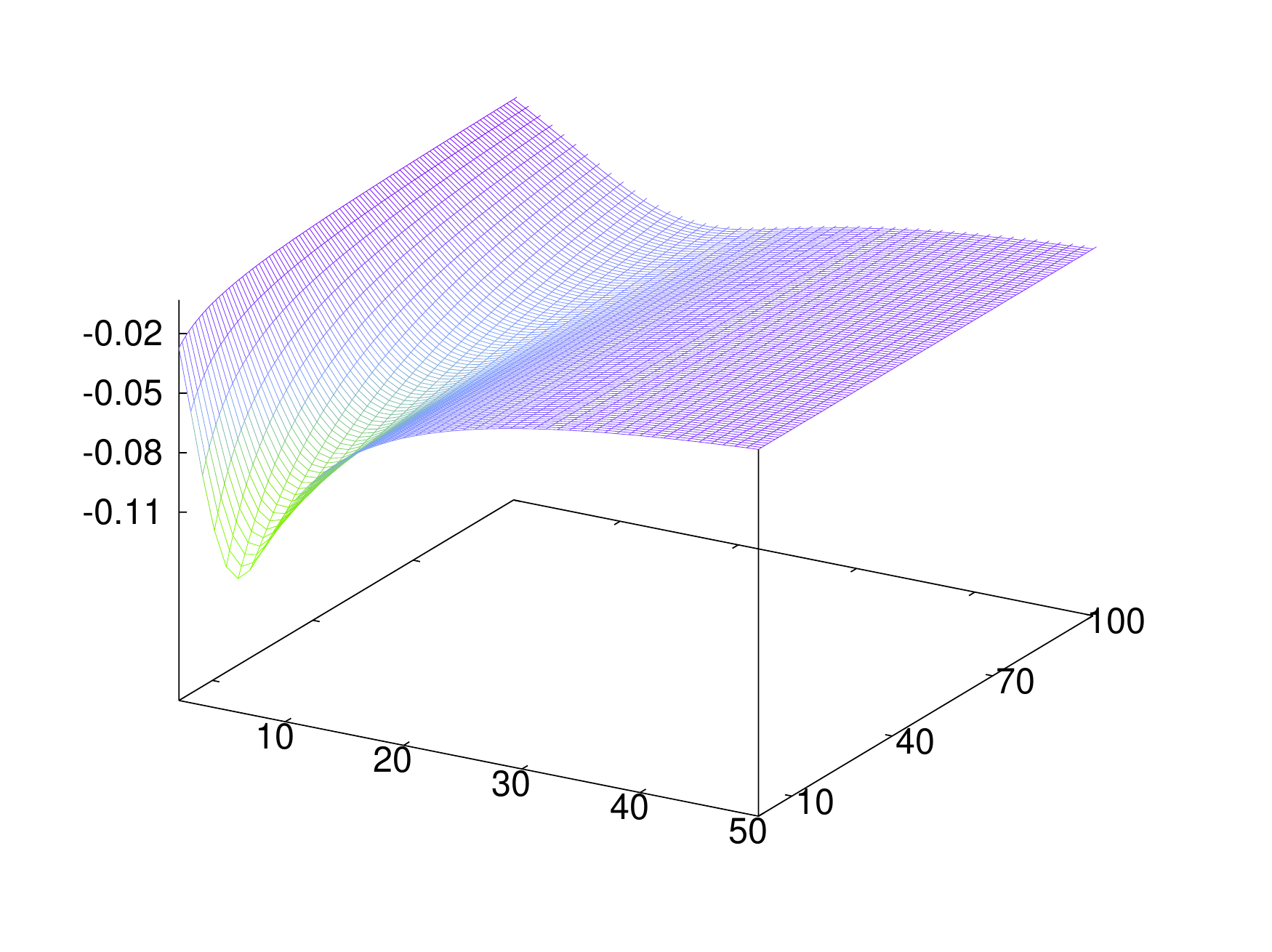}
\caption{Short range ($v^{(s)}$) and long range  ($v^{(l)}$) parts of an attractive Coulomb
potential. The  parameters are $Z=-1$, $x_{0}=4$, $y_{0}=12$ and $\nu=2.1$. }
\label{v1}
\end{figure}

\begin{figure}
\includegraphics[width=0.5\textwidth]{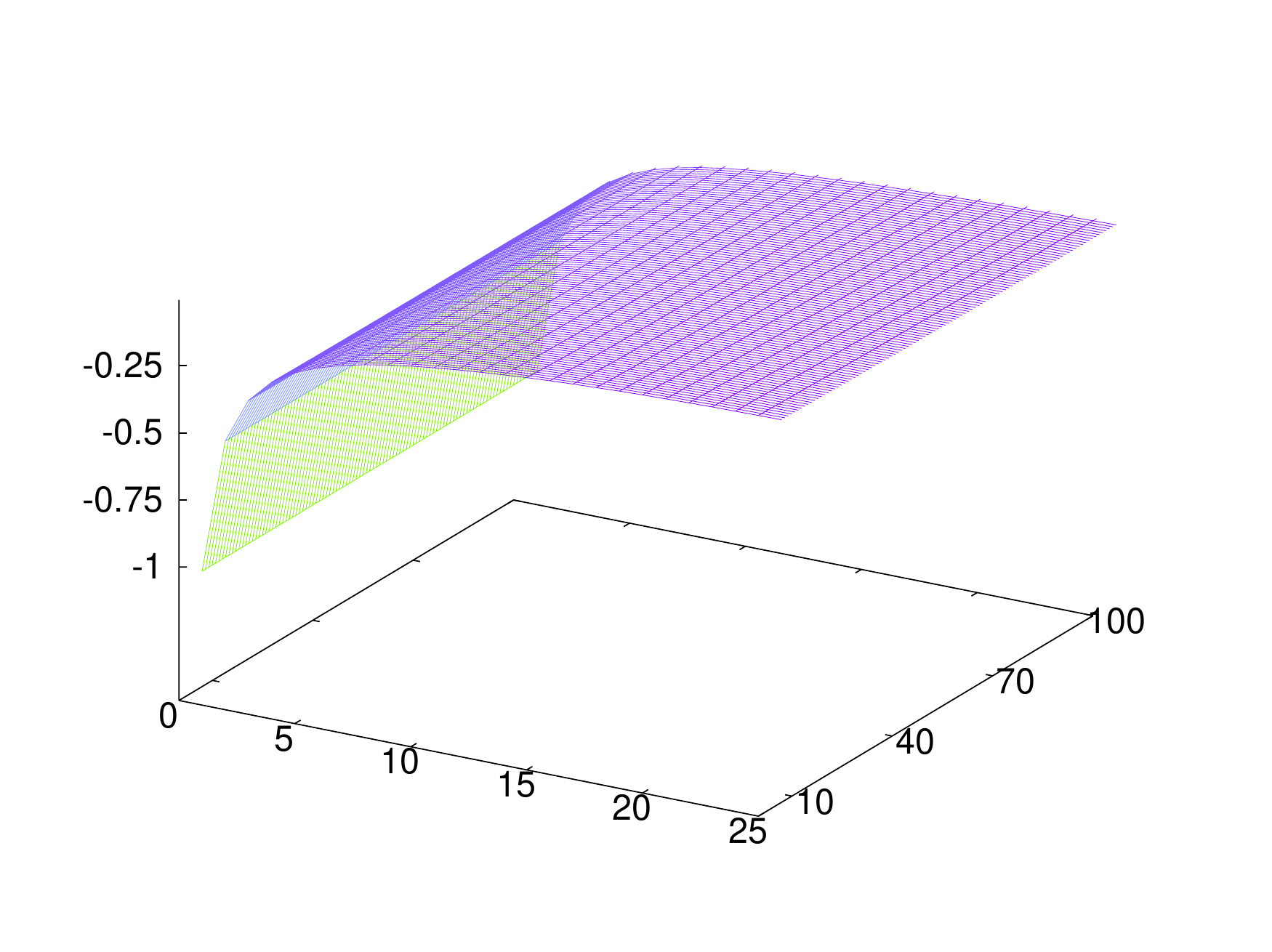}\includegraphics[width=0.5\textwidth]{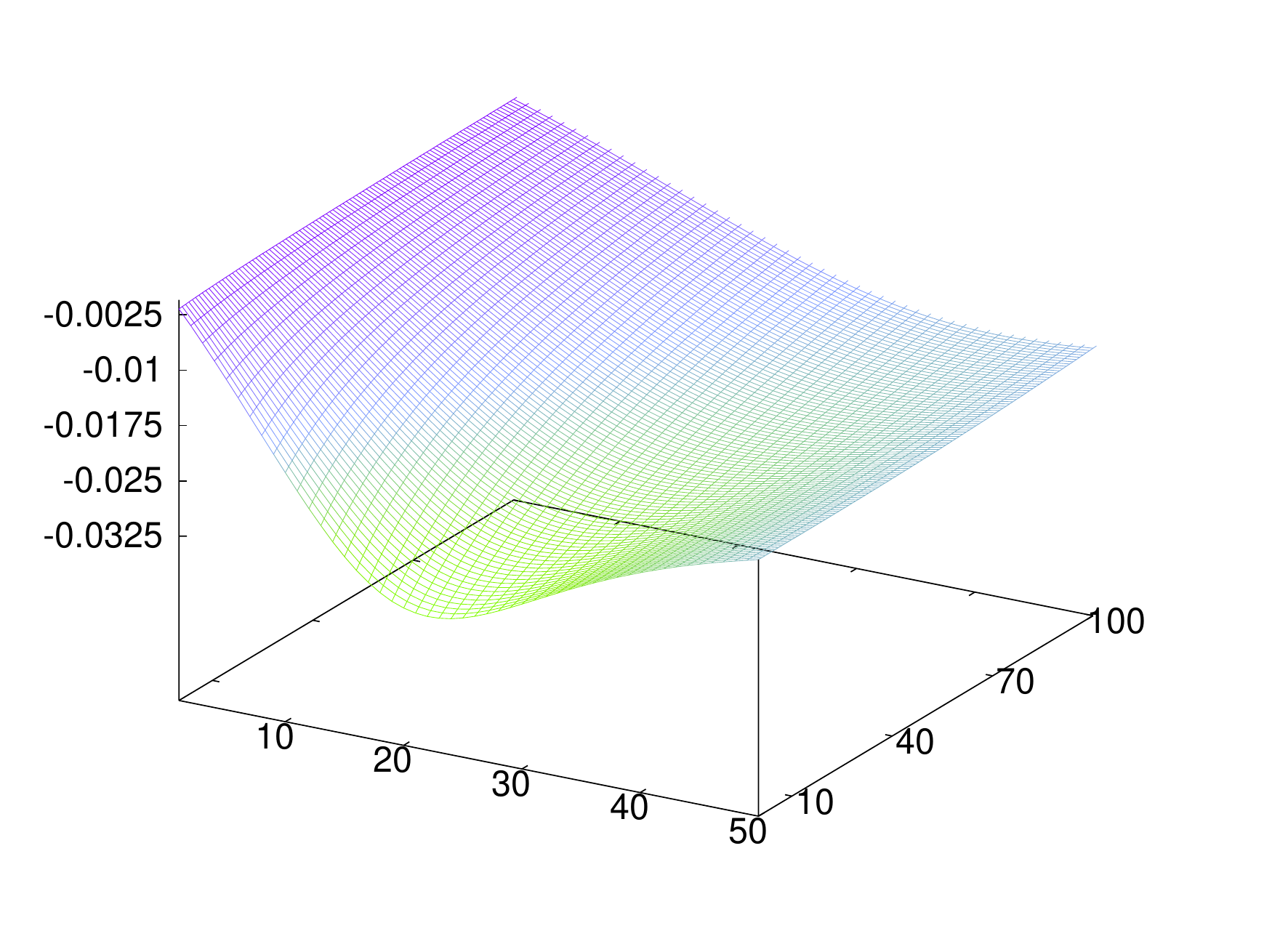}
\caption{Short range ($v^{(s)}$) and long range  ($v^{(l)}$) parts of an attractive Coulomb
potential. The  parameters are $Z=-1$, $x_{0}=15$, $y_{0}=35$ and $\nu=2.1$. }
\label{v2}
\end{figure}

Therefore we split the Coulomb potential  in 
the three-body configuration space into
short-range and long-range parts
\begin{equation}
v^C_\alpha =v^{(s)}_\alpha +v^{(l)}_\alpha.
\label{pot}
\end{equation}
The splitting is carried out with the help of a splitting function 
$\zeta_\alpha$,
\begin{eqnarray}
v^{(s)}_\alpha (x_\alpha,y_\alpha) & = & v^C_\alpha(x_\alpha) 
\zeta_\alpha (x_\alpha,y_\alpha),
\\
v^{(l)}_\alpha (x_\alpha,y_\alpha) & = & v^C_\alpha(x_\alpha) 
\left[1- \zeta_\alpha (x_\alpha,y_\alpha) \right].
\label{potl}
\end{eqnarray}
The function  $\zeta_\alpha$  
vanishes asymptotically within the three-body sector,
where $x_\alpha\sim y_\alpha \to \infty$, and approaches $1$ in 
the two-body asymptotic region
$\Omega_\alpha$, where $x_\alpha << y_\alpha \to \infty$. 
As a result, in the three-body sector, the short-range potential $v^{(s)}_\alpha$
vanishes and long-range potential $v^{(l)}_\alpha$ approaches $v^{C}_\alpha$.
In practice, the functional form
\begin{equation}
\zeta_\alpha (x_\alpha,y_\alpha) =  
2/\left\{1+ \exp 
\left[ {(x_\alpha/x_{0})^\nu}/{(1+y_\alpha/y_{0})} 
\right] \right\}
\label{zeta}
\end{equation}
is used. Typical shapes for $v^{(s)}$ and $v^{(l)}$ 
are shown in Figures \ref{v1}  and \ref{v2}. In fact, these parameters were adopted in our $e-Ps$ calculations. We can see
that $v^{(l)}$ is a valley which opens up as 
$y_{\alpha}$ goes to  infinity and becoming shallower and shallower. Finally, in the $y_{\alpha} \to \infty$
limit there is no two-body bound state in $x_{\alpha}$. 

The Coulomb potential $v_3^C$ is repulsive. 
So, it does not support bound states and there are no two-body channels associated with this 
fragmentation. Consequently, the entire $v_3^C$ can be considered as a long-range potential. Then
the long-range Hamiltonian is defined as
\begin{equation}
H^{(l)} = H^0 + v_1^{(l)}+ v_2^{(l)}+ v_3^{C},
\label{hl}
\end{equation}
and the three-body Hamiltonian looks like an ordinary three-body Hamiltonian with only two
short range interactions
\begin{equation}
H = H^{(l)} + v_1^{(s)}+ v_2^{(s)}.
\label{hll}
\end{equation}
So, we split the wave function into two components only
\begin{equation}\label{psi12}
|\Psi \rangle =| \psi_1 \rangle + | \psi_2 \rangle.
\end{equation}
Then for components, we have the set of  Faddeev equations,
\begin{eqnarray}
(E-H_{1}^{(l)}) | \psi_1 \rangle &= &  v_1^{(s)} 
| \psi_2  \rangle \nonumber \\ 
(E-H_{2}^{(l)}) | \psi_2 \rangle &= &  v_2^{(s)} | \psi_1 \rangle,
\label{fm2c2d}
\end{eqnarray}
where 
\begin{equation}\label{halpha}
H_\alpha^{(l)}=H^{(l)}+v_\alpha^{(s)}.
\end{equation} 
By adding these two equations  we recover the original 
Schr\"odinger equation. So, the Faddeev procedure is a clever way of solving the quantum mechanical Schr\"odinger equation.
We can write these differential equations into an integral equation form
\begin{eqnarray}
| \psi_1 \rangle &= & G_1^{(l)} (E) v_1^{(s)} 
| \psi_2  \rangle  \nonumber \\  
| \psi_2 \rangle &= & 
G_2^{(l)}(E) v_2^{(s)} | \psi_1 \rangle,
\label{fm2c2}
\end{eqnarray}
where $G_\alpha^{(l)}(E)=(E-H_{\alpha}^{(l)})^{-1}$.  
With Merkuriev's  procedure Faddeev's aim is achieved for the Coulomb potential as well. Now each component describes only 
one kind of asymptotic fragmentation.

If  particles $1$ and $2$ are identical particles,  the Faddeev components 
$| \psi_1 \rangle$ and $| \psi_2 \rangle$, in their own natural Jacobi
coordinates, must have the same functional forms
\begin{equation}
\langle x_1 y_1 | \psi_1 \rangle = \langle x_2 y_2 | \psi_2 \rangle.
\end{equation}
On the other hand, by interchanging particles $1$ and $2$, we have
\begin{equation}
{\mathcal P_{1 2}} | \psi_1 \rangle = p | \psi_2 \rangle,
\end{equation}
where  $p=\pm 1$, depending the total spin of the two identical particles.
So, $\psi_{1}$ and $\psi_{2}$ are not independent, and to determine one of them we need only one equation 
\begin{equation} \label{fmp}
| \psi_{1} \rangle =   G_1^{(l)} v_1^{(s)} p {\mathcal P_{1 2}} 
| \psi_{1} \rangle.
\end{equation}
As we can see, we can easily incorporate the identity of particles into the Faddeev formalism, and this even leads to 
a considerable simplification of the equations.

\subsection{Solution method}

In order that we can solve the Faddeev-Merkuriev integral equations we represent them in Coulomb--Sturmian (CS) basis.
The CS functions are given by
\begin{equation}
\langle r\vert n l ;b  \rangle 
= \sqrt{ \frac{n!}{(n+2l+1)!} } \ 
\exp(-b r) (2b r)^{l+1} L_n^{(2l+1)}(2b r)\ , 
\label{csf}
\end{equation}
where  $L$ denotes the Laguerre polynomials,  $l$ is angular momentum, $n$ is the radial 
quantum number and $b$ is a parameter.
With 
$\langle r\vert \widetilde{n l;b} \rangle \equiv \langle r\vert n l;b \rangle/r$, 
the orthogonality and completeness relations take the forms
\begin{equation} \label{orto}
\langle \widetilde{ n'l;b }  \vert nl;b \rangle 
= \langle n'l;b  \vert \widetilde{ n l;b }  \rangle = \delta_{n n'}
\end{equation}
and 
\begin{equation} \label{complet}
{\bf 1} = \lim_{N\to \infty} \sum_{n=0}^N \vert  \widetilde{ n l;b} \rangle 
\langle n l;b \vert 
= \lim_{N\to \infty} \sum_{n=0}^N  \vert  { n l;b} \rangle 
\langle \widetilde{n l;b}  \vert \ .
\end{equation}

The three-body Hilbert space is a direct product of two-body
Hilbert spaces, so, as a basis, we may take the angular-momentum-coupled direct product of the two-body bases, 
\begin{equation}
| n \nu  l \lambda ; b \rangle_\alpha =
 | n  l ; b \rangle_\alpha \otimes |
\nu \lambda ;b \rangle_\alpha, \ \ \ \ (n,\nu=0,1,2,\ldots),
\label{cs3}
\end{equation}
where $| n  l ;b \rangle_\alpha$ and $|\nu \lambda ;b \rangle_\alpha$ 
are associated
with the coordinates $x_\alpha$ and $y_\alpha$, respectively.
With this basis, the completeness relation
takes the form (with angular momentum summation implicitly included)
\begin{equation}
{\bf 1} =\lim\limits_{N\to\infty} \sum_{n,\nu=0}^N |
 \widetilde{n \nu l \lambda ;b } \rangle_\alpha \;\mbox{}_\alpha\langle
{n \nu l \lambda ;b } | =
\lim\limits_{N\to\infty} {\bf 1}^{N}_\alpha.
\end{equation}

We insert a unit operator into the Faddeev equations (\ref{fm2c2}) 
\begin{eqnarray}
| \psi_1 \rangle &= & \lim_{N\to\infty } G_1^{(l)} (E) {\bf 1}^{N}_{1 } v_1^{(s)} 
 {\bf 1}^{N}_{2 } | \psi_2  \rangle  \nonumber \\  
| \psi_2 \rangle &= &  \lim_{N\to\infty } G_2^{(l)}(E)  {\bf 1}^{N}_{2 } v_2^{(s)} {\bf 1}^{N}_{1 } | \psi_1 \rangle,
\label{fm2c22}
\end{eqnarray}
and keep $N$ finite. This amounts to  approximating $v_{\alpha}^{(s)}$  in the three-body
Hilbert space  by a separable form
\begin{eqnarray}
v_{\alpha}^{(s)}  & & =  \lim_{N\to\infty} 
{\bf 1}^{N}_{\alpha} v_\alpha^{(s)}  {\bf 1}^{N}_\beta  \approx 
{\bf 1}^{N}_\alpha v_\alpha^{(s)}  {\bf 1}^{N}_\beta  \approx \nonumber \\ 
& & 
\sum_{n,\nu ,n', \nu'=0}^N
|\widetilde{n\nu l \lambda,b}\rangle_\alpha \; \underline{v}_{\alpha \beta}^{(s)}
\;\mbox{}_\beta \langle \widetilde{n' \nu' l' \lambda';b}|,  \label{sepfe}
\end{eqnarray}
where $\underline{v}_{\alpha \beta}^{(s)}=\mbox{}_\alpha \langle n\nu l \lambda; b |
v_\alpha^{(s)}  |n' \nu' l' \lambda' ; b \rangle_\beta$.\
In general, we can calculate these matrix elements  numerically.
The completeness of the CS basis guarantees the convergence of the expansion
with increasing $N$ and angular momentum channels.

This approximation turns the homogeneous
Faddeev--Merkuriev equation
into  a matrix equation for the component vector
\begin{eqnarray}\label{fm2comp}
\underline{\psi}_1  &= & \underline{G}_1^{(l)} (E) \underline{ v}_{1 2}^{(s)} 
  \underline{\psi}_2   \nonumber \\  
\underline{ \psi}_2  &= &  \underline{G}_2^{(l)}(E)  \underline{ v}_{2 1}^{(s)} \underline{\psi}_1,
\label{fm2c24}
\end{eqnarray}
where
\begin{equation}
\underline{G}_{\alpha}^{(l)}=\mbox{}_\alpha \langle 
\widetilde{n\nu l\lambda;b } |G_\alpha^{(l)}|\widetilde{n' \nu' l' \lambda';b } \rangle_\alpha.
\end{equation}

The Green's operator  $\underline{G}_\alpha^{(l)}$ is too complicated for a direct evaluation.\
However, in the Faddeev-Merkuriev equation it generates only $\alpha$-type asymptotic configurations where 
particles $\beta$ and $\gamma$ form bound or scattering states. Therefore in this region of the three-body configuration space
$G_{\alpha}^{(l)}$ can be  linked to a simpler Green's operator
\begin{equation}
G_{\alpha}^{(l)}(z)=\tilde{G}_\alpha(z) + \tilde{G}_\alpha (z) U_\alpha G_\alpha^{(l)}(z).
\label{LSass}
\end{equation}
where 
$\tilde{G}_{\alpha} (z)=(z-\tilde{H}_{\alpha})^{-1}$  and $U_\alpha=H_{\alpha}^{(l)}-\tilde{H}_{\alpha} $
with
\begin{equation}
\tilde{H}_{\alpha}=H^{0}+v_{\alpha}^{C}+u_{\alpha}^{(l)}.
\label{htilde}
\end{equation}
Here $u_{\alpha}^{(l)}(y_{\alpha})=Z_{\alpha} (Z_{\beta} +Z_{\gamma})e^{2}/y_{\alpha}$. This way $U_{\alpha}$ is of short range type, 
and can be approximated on the CS basis as before.

In our Jacobi coordinates, the three-particle kinetic energy
can be written  as a sum of two-particle free Hamiltonians 
\begin{equation}
H^0=h_{x_\alpha}^0+h_{y_\alpha}^0.
\end{equation}
Thus the Hamiltonian $\widetilde{H}_\alpha$ of Eq.\ (\ref{htilde}) 
appears as a sum of two two-body Hamiltonians acting on different coordinates 
\begin{equation}
\widetilde{H}_\alpha =h_{x_\alpha}+h_{y_\alpha},
\end{equation}
where $h_{x_\alpha}=h_{x_\alpha}^0+v_\alpha^C(x _\alpha)$ and $h_{y_\alpha}=h_{y_\alpha}^0+u_{\alpha}^{(l)}(y_{\alpha})$. 
So,  $\widetilde{G}_\alpha$
is a resolvent of the sum of two commuting Hamiltonians $h_{x_{\alpha}}$ and $h_{y_{\alpha}}$.
Such  resolvents can be expressed as a 
convolution integral of two-body Green's operators
\begin{eqnarray}
\widetilde{G}_\alpha (z)&=& (z- h_{y_{\alpha}}-h_{x_{\alpha}})^{-1}  \nonumber \\
&=& \frac 1{2\pi {i}}\oint_{\cal C}  dz' \,(z-h_{y_{\alpha}}-z')^{-1} \;(z'-h_{x_{\alpha}})^{-1}   \nonumber \\
&=& \frac 1{2\pi {i}}\oint_{\cal C}  dz' \,g_{y_\alpha}(z-z')\;g_{x_\alpha}(z'),
\label{contourint}
\end{eqnarray}
where
$g_{x_\alpha}(z)=(z-h_{x_\alpha})^{-1}$  and
$g_{y_\alpha}(z)=(z-h_{y_\alpha})^{-1}$.
The contour $\cal C$ should be taken in a counterclockwise direction
around the singularities of $g_{x_\alpha}$
such that $g_{y_\alpha}$ is analytic on the domain encircled
by $\cal C$. So, to calculate the matrix elements $\widetilde{\underline{G}}_\alpha$, we need to calculate a contour 
integral of the two-body Green's matrices $\underline{g}_{y_\alpha}$ and $\underline{g}_{x_\alpha}$. Those two-body Coulomb
Green's matrix elements can be calculated analytically for complex energies by continued fractions \cite{dhp}. This is an exact representation of 
$g_{x_\alpha}$ and $g_{y_\alpha}$, consequently the thresholds are at the exact locations with the proper Coulomb degeneracy.

\begin{figure}[htbp]
\includegraphics[width=0.8\textwidth]{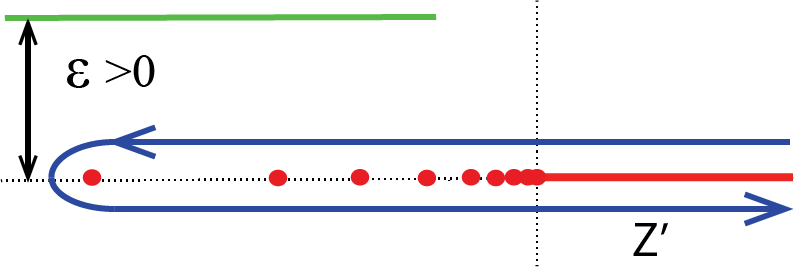}
\caption{The analytic structure of $g_{y_1}(E_{r}+{\mathrm{i}}\varepsilon-z')
g_{x_1}(z')$ as a function of $z'$ with $\varepsilon>0$. The operator
$g_{x_1}(z')$ has a branch-cut on the $[0,\infty)$ interval and accumulation of infinitely many
bound states at zero energy, while
$g_{y_1}(E+{\mathrm{i}}\varepsilon-z')$ has a branch-cut on the 
$(-\infty,E+{\mathrm{i}}\varepsilon]$ interval.
The contour $C$ encircles the spectrum of
$g_{x_1}$ and avoids the singularities of $g_{y_{1}}$. }
\label{fig1}
\end{figure}

\begin{figure}[htbp]
\includegraphics[width=0.8\textwidth]{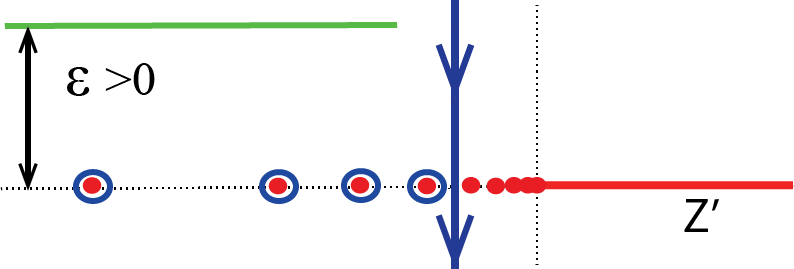}
\caption{The contour of Fig.\ \ref{fig1} is
deformed analytically such that it encircles the low-lying bound-state poles of $g_{x_1}$ and the other part is taken 
along an imaginary direction. }
\label{fig2}
\end{figure}

\begin{figure}[htbp]
\includegraphics[width=0.8\textwidth]{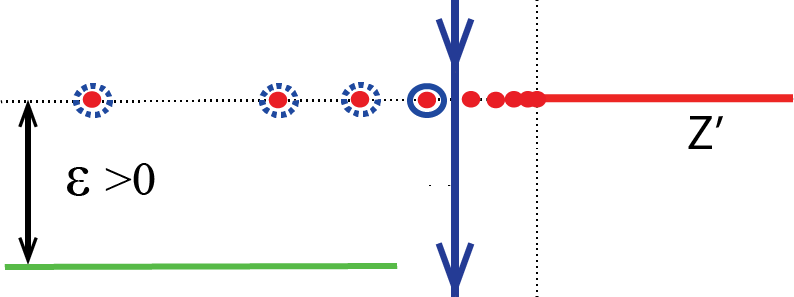}
\caption{Analytic continuation to   $\varepsilon = - \Gamma/2$.  The low-lying poles of $g_{x_{1}}$ submerge onto the 
second Riemann sheet of $g_{y_{1}}$, and they are denoted by dotted contour.}
\label{fig4}
\end{figure}

In this work we calculate the negative energy resonances of the $e-Ps$ three-body system. We need to solve (\ref{fmp}) such that
in $x_{1}$ we have the $e^{-}-e^{+}$ pair.
So, $g_{x_{1}}$ is a Coulomb Green's operator with a branch-cut on the  $[0,\infty)$ interval 
and accumulation of infinitely many bound states at zero energy. On the other hand $u_{1}^{(l)}(y_{1})$ is absent and  $g_{y_{1}}$ is a free Green's operator with 
branch-cut singularity on the $[0,\infty)$ interval. The resonances are at $E=E_{r}- \mathrm{i} \Gamma/2$. 
First, we need to formulate
$\widetilde{G}_1 (E)=  \widetilde{G}_1 (E_{r} +{\mathrm{i}}\varepsilon)$, with $\varepsilon > 0$, then 
we need continue analytically to $E=E_{r}- \mathrm{i} \Gamma/2$.
For this purpose we take the contour of Fig.\ \ref{fig1}. 
With $\varepsilon>0$ 
the singularities of $g_{x_1}$ and $g_{y_1}$ are well separated and the contour encircles 
the spectrum of $g_{x_1}$ without touching 
the singularities of $g_{y_1}$.  Then we change the contour analytically as shown in Fig.\ \ref{fig2}. 
The contour encircles some low-lying singularities of $g_{x_{1}}$ resulting in its residue, while
the other part of the contour is deformed to an integration along a straight line parallel to the imaginary axis. 
Now, we can take the $\varepsilon \to - \Gamma/2$ transition. By doing so, the poles of $g_{x_{1}}$ submerge 
into the second Riemann sheet of $g_{y_1}$  but the contour stays away from the singularities of $g_{y_1}$ (Fig.~\ref{fig4}).

In calculating the three-body Coulomb Green's matrix $\widetilde{G}_{\alpha}$ the mathematical condition for the integral in Eq.\ (\ref{contourint}) is that the contour $\cal C$ should encircle the spectrum of one of the two-body Green's operators 
without incorporating the spectrum of the other. In Refs.\ \cite{prl,zsolti} the contour was taken such a way that it encircled the singularities of $g_{y}$. However, for resonant-state energies,
the bound-state poles of $g_{x}$ penetrate into the continuous spectrum of $g_{y}$. Then to meet the requirement for the contour $\cal C$, the path around the spectrum of $g_{y}$ had to 
be taken in such a way that it descends down into the unphysical Riemann sheet. But, the integration on the unphysical sheet is rough, 
the Green's matrix exhibits violent changes,
and this is getting even worse for broader resonances as the contour dives deeper into the second sheet. A singularity is always very prominent, 
so this numerical inaccuracy does not eliminate the resonance poles and does not mask 
the whole phenomenon, but it makes the identification of 
individual resonances, especially the broad ones, less trustworthy. The contour adopted here avoids this pitfall. No integration goes on the 
unphysical sheet, the path of integration is far away from any singularities, so we get very reliable results with just a few integration points.

\section{Results}

We calculate the $L=0$
resonances of the $e-Ps$ system.  We used atomic units throughout. We have to select $l$ and $\lambda$ such that 
$l-\lambda=0$.
The two electrons are identical and the two spins can be coupled either to $S=0$ or $S=1$. 
The $S=0$ state is antisymmetric with respect to the exchange of the spin coordinates, and consequently it should be symmetric  
with respect to exchange the electron spatial coordinates. 
So, if $S=0$ we have $p=1$ in Eq.\ (\ref{fmp}),  while if $S=1$  we have $p=-1$. 

We have two parameters to vary in the calculations. One is the  scale parameter $b$. We found a good stability 
in our results with $b= 0.25$. The other parameter is $N$, the maximal radial quantum number in the expansion of 
the potentials in each angular momentum channel
and in $x$ and $y$ coordinates.  
We can see that while the narrow resonances are very stable with increasing $N$, the broad 
resonances  are not so. This is understandable since the broad resonances are 
always hard to calculate. But this inaccuracy does not change the whole picture. Individual resonances may vary a little bit 
with increasing 
$N$ and with changing the parameters $x_{0}$ and $y_{0}$, 
but the series of resonances originating from the threshold are there.  We found $N=32$ big enough for stable results.

\begin{figure}[htbp]
\includegraphics[width=0.5\textwidth]{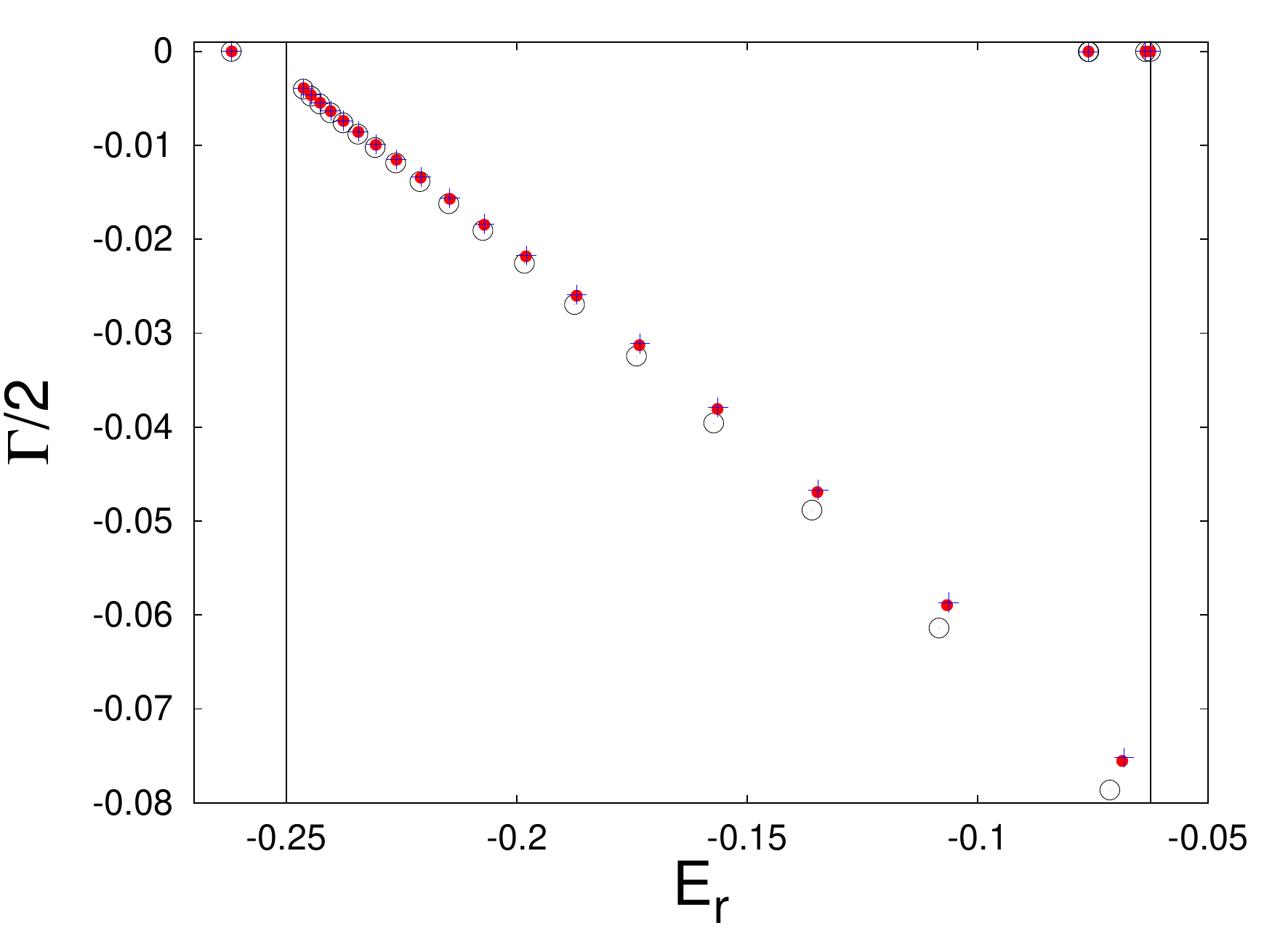}\includegraphics[width=0.5\textwidth]{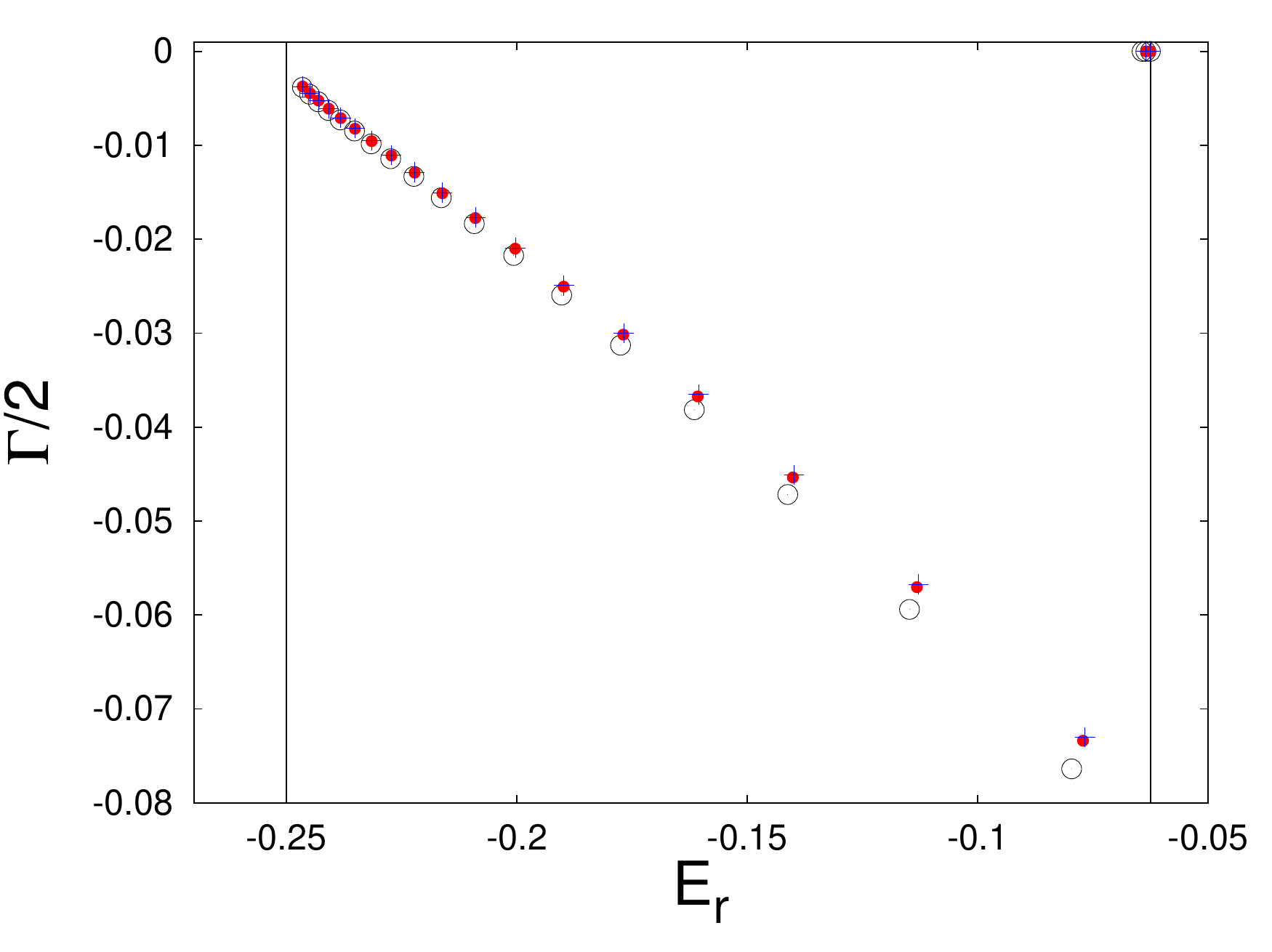}
\caption{ $^1S$ and $^3S$ resonances of the $e-Ps$ system between the $n=1$ and $n=2$ thresholds. 
Black circles indicate data using cut-off parameters $x_0 = 4$, $y_0=12$, red dots indicate data using cut-off 
parameters $x_0 = 8$, $y_0 = 24$,  and blue + symbols indicate data using cut-off parameters $x_0 = 8$, $y_0 = 12$. Thresholds indicated with vertical bars.}
\label{res12}
\end{figure}

\begin{figure}[htbp]
\includegraphics[width=0.5\textwidth]{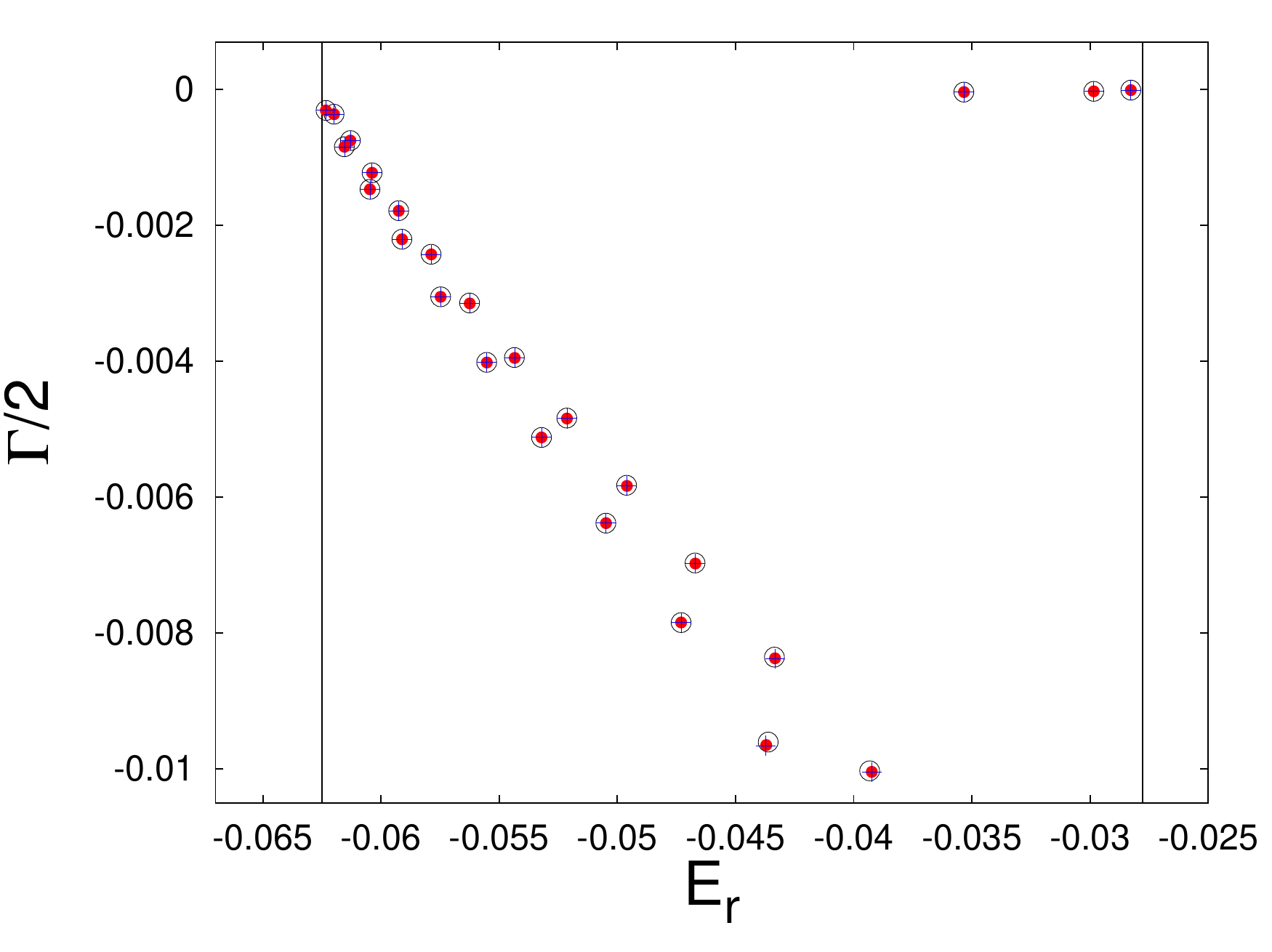}\includegraphics[width=0.5\textwidth]{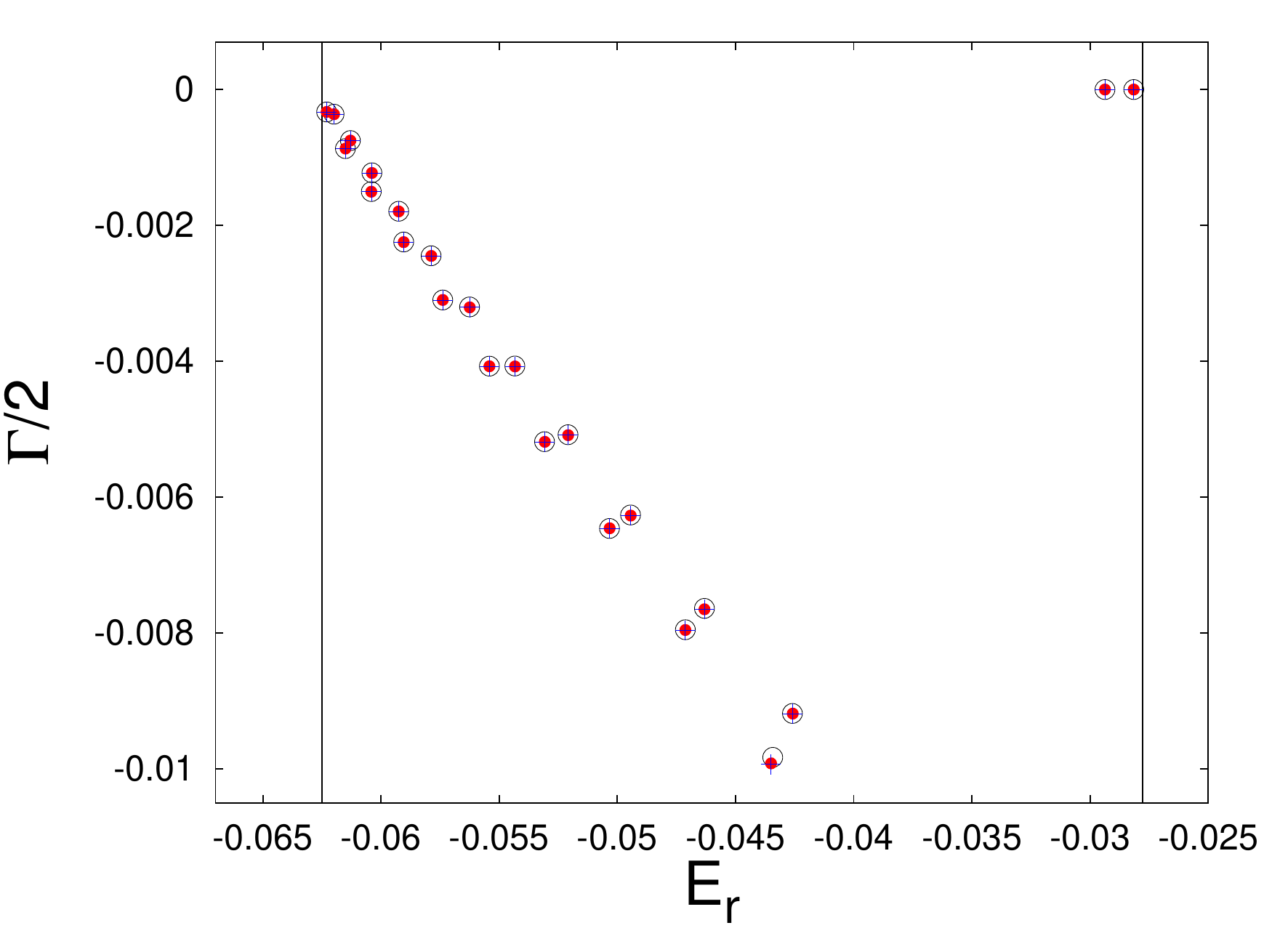}
\caption{$^1S$ and $^3S$ resonances of the $e-Ps$ system between the $n=2$ and $n=3$ thresholds. 
Black circles indicate data using cut-off parameters $x_0 =12$, $y_0=30$, red dots indicate data using cut-off 
parameters $x_0 = 15$, $y_0 = 35$, and blue + symbols indicate data using cut-off parameters $x_0 = 15$, $y_0 = 30$. Thresholds indicated with vertical bars. }
\label{res23}
\end{figure}

Figures (\ref{res12}) and (\ref{res23})   show the resonances between various thresholds using three cut-off parameters.
False resonances may occur in the 
Faddeev-Merkuriev method. They are associated with the possible bound states of $H^{(l)}$.  But, by taking $x_{0}$  about the same size as the 
two-body subsystem in $x$, and varying $(x_{0},y_{0})$, we can avoid them. We can see that the broad resonances 
line up along a straight line with increasing spacings. It seems that at higher thresholds we have more lines.

\section{Summary and Interpretation}

In this work we re-examined the broad resonances lining up around thresholds. We solved the Faddeev-Merkuriev integral 
equations by adopting a Coulomb-Sturmian based separable expansion approach on the potential in the three-body
configuration space. This method approximates only the asymptotically irrelevant short range potentials. The asymptotically relevant parts are 
kept in the Green's operator $\widetilde{G}_\alpha$, and its CS matrix elements have been evaluated as a complex convolution integral
of the two-body Green's matrices. The adopted contour makes the calculation of $\widetilde{G}_\alpha$ numerically exact, even for very broad resonances,
and ensures that all the thresholds are at the right location. The only real approximation is that the basis in each coordinates are truncated to a finite $N$, but 
we found $N=32$ big enough for very reliable results.

We found that some of those resonances lie along a straight line. From Fig.\ \ref{res12} we can see that for those resonances the ratio 
$\epsilon_{m}/\Delta \epsilon_{m} \approx 1.23 \pm 0.03$, 
where $\epsilon$ is the center of mass energy measured form the  thresholds. They must have a common origin. 

It appears that these series of resonances have not been reproduced by other methods. Careful examination and comparison with previous resonance
calculations revealed that long-lived resonances were calculated previously along with Feshbach resonances. They are called shape resonances
in the atomic physics community. As early as 1962-1963 in the work of Gailitis and Damburg \cite{gailitis} their presence as T-matrix oscillations 
above thresholds were calculated.

Hu and Caballero \cite{hc} carried out a six open-channel high precision calculation of the $e^{+}+H(n=2)$ scattering system using the Faddeev-Merkuriev
differential equations. The calculation involves no intermediate approximations nor truncation of any kind. 
The numerical method solved a half million coupled linear equations. All scattering properties were calculated in the range of energy just above the 
$Ps(n=2)$ formation threshold. The results display singularities of the $K$-matrix ($\tan \delta$) (the phase shift jumps by $\pi$), the cross section maxima and the six channel wave amplitudes.
All display three resonances within a cutoff distance of $1000 a_{0}$, where $a_{0}$ is the Bohr radius, in channels $p+Ps(n=2,l=0)$ and
$p+Ps(n=2,l=1)$. 
These calculations revealed a previously unknown, but relatively simple formation mechanism for these kind of resonances. 

For the $Ps(n=2)$ target, the Coulomb degeneracy allows the incoming proton to induce a first order constant electric dipole moment $\mu_{1}$ on the target,
known as the first order Stark effect. 
By analyzing the channel wave functions in resonance channels $p+Ps(n=2,l=0)$ and $p+Ps(n=2,l=1)$ Hu and Caballero  \cite{hc} found that 
the resonant conditions arise when the center-of-mass energy of the proton satisfy the simple relations
\begin{equation}
\epsilon_{m}= m \mu_{1}/y_{m}^{2},\ \ \ \ \epsilon_{m}/\Delta \epsilon_{m} = \mbox{constant},  
\end{equation}
where $m=1,2, \ldots$ integer, and $y_{m}$ is the location (Jacobi coordinate) of the proton at the resonance.

The Stark effect is a universal phenomenon. Furthermore, if the target is not degenerate the second order Stark effect takes place. 
The induced electric dipole moment is $\mu_{1}=\alpha /y^{2}$, where $\alpha$ is the second
order polarizability. Then the resonant condition becomes 
\begin{equation}
\epsilon_{m}= m \mu_{1}/y_{m}^{2}= m \alpha/y_{m}^{4}, \ \ \ \ \ m = 1,2,\cdots.
\end{equation}

These explain our observations that the resonances are lining up from the thresholds with increasing spacing. Also, in the energy region between $n=2$ and $n=3$ thresholds
the induced electric dipole moment $\mu$ is different for $l=0$ and $l=1$, so we observe some splitting of the lines as well.

{\bf Author Contributions:} This method of calculating three-body resonances has been designed by Z.\ Papp, the calculations were carried out by D.\ Diaz, and the 
interpretation has been provided by C.-Y.\ Hu.

{\bf Conflict of Interest:} The author declare no conflict of interest.


\end{document}